\documentstyle[12pt,amssymb]{article}
\begin{document}
\newcommand{\iint}{{\int\hskip-3mm\int}}
\newcommand{\dddot}[1]{{\mathop{#1}\limits^{\vbox to 0pt{\kern 0pt
 \hbox{.{\kern-0.25mm}.{\kern-0.25mm}.}\vss}}}}

\renewcommand{\theequation}{\thesection.\arabic{equation}}
\let\ssection=\section
\renewcommand{\section}{\setcounter{equation}{0}\ssection}

\newcommand{\be}{\begin{equation}}
\newcommand{\ee}{\end{equation}}
\newcommand{\bes}{\begin{eqnarray}}
\newcommand{\ees}{\end{eqnarray}}

\newcommand{\qed}{{\hbox{$\  \Box$}}}

\newtheorem{theorem}{Theorem}[section]
\newtheorem{lemma}[theorem]{Lemma}
\newtheorem{proposition}[theorem]{Proposition}
\newtheorem{definition}[theorem]{Definition}

\newcommand{\dt}[1]{\dot#1}
\newcommand{\ddt}[1]{\ddot#1}
\newcommand{\dddt}[1]{\dddot#1}

\newcommand{\nl}{\nonumber\\}
\newcommand{\bl}{&&\qquad }
\newcommand{\ab}{\allowbreak}
\renewcommand{\/}{\over}
\renewcommand{\d}{\partial}
\newcommand{\eps}{\epsilon}
\newcommand{\dlt}{\delta}
\newcommand{\th}{\theta}
\newcommand{\al}{\alpha}
\newcommand{\si}{\sigma}
\newcommand{\om}{\omega}
\newcommand{\la}{\lambda}
\newcommand{\ka}{\kappa}
\newcommand{\Dl}{{\Delta}}
\newcommand{\txi}{{\tilde\xi}}
\newcommand{\teta}{{\tilde\eta}}
\newcommand{\vx}{\vec x}
\newcommand{\vq}{\vec q}
\newcommand{\emt}{\e^{imt}}
\newcommand{\thb}{{\bar\theta}}
\newcommand{\gb}{{\bar g}}

\newcommand{\no}[1]{{:\!#1\!:\ }}

\newcommand{\mum}{{\mu_1..\mu_m}}
\newcommand{\mun}{{\mu_1..\mu_n}}
\newcommand{\nun}{{\nu_1..\nu_n}}
\newcommand{\sumnu}[1]{\nu_1+..+\nu_{#1}}
\newcommand{\dx}{{d^{N+M+1}x}}
\newcommand{\xmu}{\xi^\mu}
\newcommand{\ynu}{\eta^\nu}
\newcommand{\zsi}{\zeta^\si}
\newcommand{\xz}{\xi^0}
\newcommand{\yz}{\eta^0}
\newcommand{\zz}{\zeta^0}
\newcommand{\dmu}{{\d_\mu}}
\newcommand{\dnu}{{\d_\nu}}
\newcommand{\dsi}{{\d_\si}}
\newcommand{\dtau}{{\d_\tau}}
\newcommand{\drho}{{\d_\rho}}
\newcommand{\qmu}{{\dt q^\mu}}
\newcommand{\qnu}{{\dt q^\nu}}
\newcommand{\qsi}{{\dt q^\si}}
\newcommand{\qtau}{{\dt q^\tau}}
\newcommand{\qrho}{{\dt q^\rho}}

\renewcommand{\L}{{\cal L}}
\newcommand{\J}{{\cal J}}
\newcommand{\Lz}{{\cal L}^0}
\newcommand{\Lxi}{\L_\xi}
\newcommand{\lxi}{\ell_\xi}
\newcommand{\Leta}{\L_\eta}
\newcommand{\Lzxi}{\Lz_\xi}
\newcommand{\Lzeta}{\Lz_\eta}

\newcommand{\nnn}{{1\/2\pi}}
\newcommand{\nnni}{{1\/2\pi i}}
\newcommand{\half}{{1\/2}}

\newcommand{\PB}{{{P.b.}}}
\newcommand{\KB}{{{K.b.}}}

\newcommand{\xy}{{\xi\eta}}
\newcommand{\xcy}{{(-)^\xy \xi\leftrightarrow\eta}}
\newcommand{\XcY}{{(-)^XY X \leftrightarrow Y}}

\newcommand{\into}{\hookrightarrow}
\newcommand{\str}{{\rm str}}
\renewcommand{\div}{{\rm \kern0.1em div\kern0.1em}}
\newcommand{\e}{{\rm e}}
\newcommand{\oj}{{\frak g}}
\newcommand{\hh}{{\frak h}}

\newcommand{\bxy}{{[\xi,\eta]}}

\newcommand{\N}{{N\!\ab+\!1}}
\newcommand{\km}[1]{{\widehat{#1}}}
\newcommand{\ext}{{\hbox{ext}}}

\newcommand{\Prim}{{\cal P}}
\newcommand{\RR}{{\Bbb R}}
\newcommand{\CC}{{\Bbb C}}
\newcommand{\ZZ}{{\Bbb Z}}
\newcommand{\NN}{{\Bbb N}}

\topmargin 1.0cm

\newpage
\vspace*{-3cm}
\pagenumbering{arabic}
\begin{flushright}
{\tt physics/9710022}
\end{flushright}
\vspace{12mm}
\begin{center}
{\huge Fock representations of non-centrally extended
super-diffeomorphism algebras}\\[14mm]
\renewcommand{\baselinestretch}{1.2}
\renewcommand{\footnotesep}{10pt}
{\large T. A. Larsson\\
}
\vspace{12mm}
{\sl Dannemoragatan 10\\
S-113 44 Stockholm, Sweden}\\
email: tal@hdd.se
\end{center}
\vspace{3mm}
\begin{abstract}
A class of Fock representations of non-central extensions
of the super-diffeomorphism algebra in $(\N|M)$ dimensions is constructed,
by superization of the paper [{\tt physics/9705040}].
The representations act on trajectories in $(N|M)$-dimensional superspace,
the extra dimension being the parameter along the trajectory.
The restrictions to various subalgebras
are considered. In particular, the centrally extended superconformal algebra 
is obtained by restriction to the contact superalgebra $K(1|1)$. 
This shows that one of the basic assumptions in superstring theory 
(the distinguished nature of the superconformal algebra) is incorrect.
\end{abstract}
\renewcommand{\baselinestretch}{1.5}

\section{Introduction}
The diffeomorphism group (and its algebra $diff(\N)$) in $\N$-dimensional 
space-time plays a crucial role in classical physics; 
suffice it to say, that local differential geometry and general relativity
may be phrased in the language of $diff(\N)$ modules and intertwiners.
However, quantum field theory is only invariant under the Poincar\'e group,
because until recently it was not known how to build projective 
$diff(\N)$ Fock modules.  The problem is that normal ordering of tensor
densities
gives rise to infinities; also, no central extension exists when $N>0$.
Recall that a module is projective if it admits a group action
up to a phase; on the Lie algebra level, this corresponds to a representation
of an abelian extension of $diff(\N)$. If the phase is local, the resulting
extension 
must transform non-trivially under the diffeomorphism algebra, i.e. it is
non-central.
Only if the phase is globally constant, the extension is equivalent to a
central one.

The first example of a projective Fock representation of $diff(\N)$ was
found by Eswara-Rao and Moody \cite{ERM94}. Analogous representations
of current algebras were previously discovered by the same group
\cite{EMY92}\cite{FM94}\cite{MEY90}.
Subsequently the present author
uncovered the geometrical meaning of their construction, and greatly
generalized
it \cite{Lar97}. It turns out that the algebra acts on trajectories
in space, the extra time dimension being the parameter along the trajectory.
The word is trajectory, not string, because we deal with one-dimensional
objects in space-time. It is remarkable that although space and time are
treated in a completely different fashion, a proper $diff(\N)$ realization on 
trajectories is obtained. Because the realization is non-linear,
normal ordering gives rise to a non-central extension of $diff(\N)$; 
the extension {\em does} distinguish between space and time.
Dzhumadildaev \cite{Dzhu96} has classified extensions 
of $diff(\N)$ by irreducible modules (i.e. tensor densities), but 
only some of the extensions that I found are covered by his theorem.
The point is that it is not sufficient to consider irreducible modules; 
interesting extensions also arise when we consider reducible but 
indecomposable modules. One example is provided by the modules defined
by the relations (\ref{Sn}) and (\ref{Rn}) below.

The super-diffeomorphism algebra $diff(\N|M)$ is the algebra of first-order
differential operators in $(\N|M)$-dimensional super space-time 
(alternative names: algebra of super vector fields $vect(\N|M)$, 
generalized Witt algebra $W(\N|M)$). The classical representations of this
algebra and its various subalgebras have been worked out in several papers
\cite{BL81}\cite{GLS97}\cite{Kac77}\cite{Lei77}\cite{Lei80}. 
For information about its bosonic counterpart, 
see e.g. \cite{Fuks86}\cite{Rud74}.
Some superalgebras possess a central extension
\cite{Ad76a}\cite{Ad76b}\cite{NS71}\cite{Ram71}; for a classification 
see \cite{GLS97}.

The purpose of the present paper is to superize the construction in
\cite{Lar97}.
It turns out that $diff(\N|M)$ has the same type of projective Fock
representations
as $diff(\N)$; superization simply amounts to a judiscious
insertion of minus signs. After some preliminaries in section 2, the main
theorem
is given in section 3. We prove that there is a classical realization on
trajectories in theorem \ref{cLxithm}, and then normal order to describe
the extension in theorem \ref{xLxithm}. Section 4 is devoted to subalgebras
of $diff(\N|M)$. Clearly, for every subalgebra $\hh\in diff(\N|M)$, 
we obtain by restriction a Fock representation of an extension of $\hh$.
In general this extension is non-central, but under special circumstances
it may reduce to a central extension (or even no extension at all).
In this way the centrally extended super-conformal algebra is obtained.
This means that the common belief that there is an exceptional algebraic 
structure underlying string theory is simply wrong. Section 5 contains a
brief statement of the corresponding representations for gauge superalgebras,
i.e. the higher-dimensional analog of the Kac-Moody superalgebra.

\section{Preliminaries}
Consider a superalgebra $\oj$ with basis $J^a$, where 
$\deg J^a \equiv \deg a = 0$ if $a$ is even (bosonic) and $\deg a = 1$ 
if $a$ is odd (fermionic). Let the symbol $(-)^a = (-1)^{\deg a}$. These
symbols satisfy an algebra modulo $2$: $a^2 = a$, $2a=0$.
A superalgebra satisfies the condition of graded skewness,
\be
[J^a, J^b] = -(-)^{ab}[J^b, J^a],
\ee
and the super-Jacobi identity,
\be
(-)^{ac}[J^a, [J^b, J^c]] + (-)^{ba}[J^b, [J^c, J^a]] 
+ (-)^{bc}[J^c, [J^a, J^a]] = 0.
\ee
In terms of structure constants, the brackets are
\be
[J^a, J^b] = if^{ab}{}_c J^c,
\ee
where $f^{ab}{}_c=0$ unless $a+b+c=0$ mod $2$, and
\bes
&&f^{ba}{}_c = -(-)^{ab} f^{ab}{}_c \nl
&&(-)^{ac} f^{bc}{}_d f^{ad}{}_e + (-)^{ab} f^{ca}{}_d f^{bd}{}_e
 + (-)^{bc} f^{ab}{}_d f^{cd}{}_e = 0.
\ees
The {\em supertrace} of a matrix $A = (A^\al_\beta)$ is 
$\str(A) = (-)^{A\al+\al} A^\al_\al$; $\str(BA) = (-)^{AB} \str(AB)$.
Henceforth, let $\oj$ be a finite-dimensional superalgebra with a graded
symmetric 
Killing metric $\dlt^{ab} \propto \str(J^a J^b)$, satisfying
$[J^a, \dlt^{bc}] = 0$, i.e.
\bes
\dlt^{ba} &=& (-)^{ab}\dlt^{ab} \nl
(-)^{ac} \dlt^{ad} f^{bc}{}_d &=& (-)^{ab} \dlt^{bd} f^{ca}{}_d
= (-)^{bc} \dlt^{cd} f^{ab}{}_d.
\label{Killing}
\ees
The associated Kac-Moody superalgebra $\km\oj$ reads
\be
[J^a(s), J^b(t)] = if^{ab}{}_c J^c(s) \dlt(s-t)
+ {k\/{2\pi i}} \dlt^{ab} \dt\dlt(s-t),
\ee
where $s, t \in S^1$.
It is the unique central extension of $map(1,\oj)$, the superalgebra of maps 
from $S^1$ to $\oj$.

The Virasoro algebra $Vir$ with central charge $c$ has the three 
equivalent forms
\bes
[L_m, L_n] &=& (n-m)L_{m+n} - {c\/12}(m^3-m) \dlt_{m+n} \nl
{[}L(s), L(t)] &=& (L(s) + L(t)) \dt\dlt(s-t)
+ {c\/ 24\pi i} (\dddt\dlt(s-t) + \dt\dlt(s-t)). \nl
{[}L_\xi, L_\eta]
&=& L_\bxy + {c\/ 24\pi i} \int dt\
 (\ddt\xi(t)\dt\eta(t) - \dt\xi(t)\eta(t)).
\label{Vir}
\ees
where $m,n\in\ZZ$, $s, t \in S^1$, $L_\xi = \int dt \xi(t) L(t)$ and
$[\xi,\eta] = \xi\dt\eta - (-)^\xy\eta\dt\xi$.
Note our sign convention, which is appropriate for lowest (as opposed to
highest)
weight representations.
The Virasoro algebra is bosonic; time possesses no useful graded
generalization.
It is compatible with $\km\oj$ in the sense that
\be
[ L(s), J^b(t)] = J^b(s) \dt\dlt(s-t).
\ee

Consider $(\N|M)$-dimensional super space-time with coordinates $x^\mu$ and 
partial derivatives $\dmu = \d/\d x^\mu$,
where greek indices $\mu, \nu = 0, 1, .., N, N+1, .. N+M$,
$\deg x^\mu = \deg \dmu = \deg \mu$.
The coordinate $t \equiv x^0$ is called {\em time}.
Further, we use latin indices $i,j = 1, .., N, N+1, .. N+M$ for
$(N|M)$-dimensional
superspace, excluding the time label $0$.
Let the first $\N$ coordinates be
bosonic (including time), and let the remaining $M$ coordinates
be fermionic. Thus, $\deg \mu = 0, \mu = 0, 1, .., N$ and $\deg \mu = 1, 
\mu = N+1, .., N+M$. Let $(-)^\mu = (-1)^{\deg \mu}$. 
The notation is consistent because time is bosonic: $(-)^0 = 1$. 
To this super space-time we associate a super-Heisenberg algebra with
generators $q^\mu$ and  $p_\nu$, satisfying
\be
[p_\nu, q^\mu] = \dlt^\mu_\nu, \qquad
[q^\mu, q^\nu] = [p_\mu, p_\nu] = 0.
\ee
Note that the brackets are graded: 
$[q^\mu,p_\nu] = (-)^{\mu\nu}[p_\nu, q^\mu]$.

The brackets of $gl(\N|M)$ are
\be
[T^\mu_\nu, T^\si_\tau] =
\dlt^\si_\nu T^\mu_\tau - (-)^{(\mu+\nu)(\si+\tau)} \dlt^\mu_\tau T^\si_\nu.
\label{glN}
\ee
The fundamental $gl(\N|M)$ representations are contravariant vectors,
covariant vectors, and densities of weight $\ka$.
\bes
[T^\mu_\nu, u^\si] &=& \dlt^\si_\nu u^\mu, \nl
{[}T^\mu_\nu, v_\tau] &=& -(-)^{(\mu+\nu)\tau} \dlt^\mu_\tau  v_\nu, \nl
{[}T^\mu_\nu, w] &=& -\ka (-)^\mu \dlt^\mu_\nu w,
\label{glfund}
\ees
respectively. The action on a general tensor density is given by tensoring.
The associated Kac-Moody algebra $\km{gl(\N|M)}$ reads
\bes
[T^\mu_\nu(s), T^\si_\tau(t)] &=& ( \dlt^\si_\nu T^\mu_\tau(s) 
 - (-)^{(\mu+\nu)(\si+\tau)} \dlt^\mu_\tau T^\si_\nu(s) )\dlt(s-t) \nl
&&- {1\/{2\pi i}} ( k_1 (-)^\mu \dlt^\mu_\tau \dlt^\si_\nu
 + k_2 (-)^{\mu+\si} \dlt^\mu_\nu \dlt^\si_\tau ) \dt\dlt(s-t) \nl
{[}L(s), T^\mu_\nu(t)] &=& T^\mu_\nu(t) \dt\dlt(s-t)
\label{kmgln}
\ees
There are two independent central charges $k_1$ and $k_2$, because
$gl(\N|M) \cong sl(\N|M)\oplus gl(1)$.

The special subalgebra $sl(\N|M)$ consists of operators satisfying
$T^\mu_\mu \equiv 0$. The fundamental representations are as
in (\ref{glfund}), except that $\ka=0$. The Kac-Moody algebra
$\km{sl(\N|M)}$ has only one independent central extension, and the brackets
are as in (\ref{kmgln}) with $k_1 = -(N+1-M)k_2$.

The super-diffeomorphism algebra $diff(\N|M)$ is the algebra of first-order 
differential operators (i.e. vector fields) in $(\N|M)$-dimensional 
super space-time.
Locally, such a vector field takes the form $\xi = \xmu(x)\dmu$. 
The brackets read
\be
[\xi, \eta] = \xmu\dmu\ynu\dnu - (-)^\xy \ynu\dnu\xmu\dmu
 = -(-)^\xy[\eta,\xi]
\ee
The {\em divergence} of a vector field is is
\be
\div\xi = (-)^{\xi\mu+\mu} \dmu\xmu.
\label{div}
\ee
$diff(\N|M)$ has generators $\Lxi$ and brackets
\be
[\Lxi,\Leta] = \L_\bxy.
\label{diffN}
\ee
The classical representations are tensor densities, 
corresponding to the following expression for $\Lxi$.
\be
\Lxi = \xmu(q) p_\mu + (-)^{(\xi+\mu)\nu+\mu} \dnu \xmu(q) T^\nu_\mu,
\label{tensors}
\ee
where the $T^\mu_\nu$ satisfy $gl(\N|M)$ (\ref{glN}). 
One shows by direct calculation that (\ref{diffN}) is satisfied.
The representations inherited from (\ref{glfund}) are
\bes
[\Lxi, \Phi^\nu(x)] &=& -\xmu(x)\dmu\Phi^\nu(x) 
 + (-)^{(\xi+\nu)\mu+\nu} \dmu\xi^\nu(x)\Phi^\mu(x), \nl
{[}\Lxi, \Phi_\nu(x)] &=& -\xmu(x)\dmu\Phi_\nu(x) 
 - (-)^{\xi\nu+\mu+\nu} \dnu\xmu(x)\Phi_\mu(x), \nl
{[}\Lxi, \Phi(x)] &=& -\xmu\dmu\Phi(x) - \ka\div\xi(x)\Phi(x),
\label{vector}
\ees
respectively. We sometimes need a alternative version of (\ref{vector}),
acting on $\Psi^\nu = (-)^\nu \Phi^\nu$ and $\Psi_\nu = (-)^\nu \Phi_\nu$.
\bes
[\Lxi, \Psi^\nu(x)] &=& -\xmu(x)\dmu\Psi^\nu(x) 
 + (-)^{(\xi+\nu+\mu)\mu} \dmu\xi^\nu(x)\Psi^\mu(x), \nl
{[}\Lxi, \Psi_\nu(x)] &=& -\xmu(x)\dmu\Psi_\nu(x) 
 - (-)^{\xi\nu} \dnu\xmu(x)\Psi_\mu(x).
\ees
The actions on a general tensor densities can be obtained 
by tensoring, if we keep the extra signs in mind. Thus, if $\Phi(x)$ and
$\Psi(x)$
are two fields (with indices suppressed),
\be
[\Lxi, \Phi(x)\otimes\Psi(x)] = [\Lxi,\Phi(x)]\otimes\Psi(x) + 
 (-)^{\xi\Phi} \Phi(x)\otimes[\Lxi,\Psi(x)]
\ee
Explicitly, the transformation law for a tensor field, with all upper indices
placed in front of the lower ones, reads
\bes
&&[\Lxi, \Phi^{\si_1...\si_p}{}_{\tau_1...\tau_q}(x)]
= -\xmu(x)\dmu\Phi^{\si_1...\si_p}{}_{\tau_1...\tau_q}(x)
 -\ka \div\xi(x) \Phi^{\si_1...\si_p}{}_{\tau_1...\tau_q}(x) \nl
\bl+ \sum_{i=1}^p (-)^{(\si_1+..+\si_i)(\mu+\si_i)+\xi\mu}
 \dmu \xi^{\si_i}(x) \Phi^{\si_1..\mu..\si_p}{}_{\tau_1...\tau_q}(x) \nl
\bl- \sum_{j=1}^q (-)^{(\si_1+..+\si_p+\tau_1+..+\tau_j+\mu)(\mu+\tau_j) +
\xi\tau_j}
 \d_{\tau_j}\xmu(x) \Phi^{\si_1...\si_p}{}_{\tau_1..\mu..\tau_q}(x).
\ees
$\deg\ \Phi^{\si_1...\si_p}{}_{\tau_1...\tau_q}(x) =
\si_1+..+\si_p+\tau_1+..+\tau_q$. 
If $\Phi^\mu$ and $\Psi_\nu$ are vector fields, 
their {\em contraction}
$\Phi^\nu(x) \Psi_\nu(x) = (-)^\nu \Psi_\nu(x)\Phi^\nu(x)$
is a scalar field. The contraction of higher tensors is defined analogously, 
but care has to be taken with signs for non-adjascent indices. 
A tensor is {\em graded symmetric} in two adjascent upper indices
if $\Phi^{\nu\mu}(x) = (-)^{\mu\nu}\Phi^{\mu\nu}(x)$ and {\em graded
skewsymmetric}
if $\Phi^{\nu\mu}(x) = -(-)^{\mu\nu}\Phi^{\mu\nu}(x)$. 
For any set of smearing functions
$ f^{\tau_1...\tau_q}{}_{\si_1...\si_p}(x)$, define
\be
\Phi(f) = \int \dx\ f^{\tau_1...\tau_q}{}_{\si_1...\si_p}(x)
\Phi^{\si_1...\si_p}{}_{\tau_1...\tau_q}(x).
\label{Phif}
\ee
Clearly, $\deg \Phi(f) = \deg\ f$. 
It transforms as $[\Lxi,\Phi(f)] = \Phi(\lxi f)$, where
\bes
&&\lxi f^{\tau_1...\tau_q}{}_{\si_1...\si_p}(x)
= \xmu(x)\dmu f^{\tau_1...\tau_q}{}_{\si_1...\si_p}(x)
+(1-\ka) \div\xi(x) f^{\tau_1...\tau_q}{}_{\si_1...\si_p}(x) \nl
\bl+ \sum_{i=1}^p (-)^{(f+\tau_1+..+\tau_q+\si_{i+1}+..+\si_p)
  (\mu+\si_i)+\xi\si_i}
  \d_{\si_i} \xmu(x) f^{\tau_1..\tau_q}{}_{\si_1..\mu..\si_p}(x)  \nl
\bl- \sum_{j=1}^q (-)^{(f+\tau_{j+1}+..+\tau_q+\mu)(\mu+\tau_j) + \xi\mu}
  \dmu\xi^{\tau_j}(x) f^{\tau_1..\mu..\tau_q}{}_{\si_1...\si_p}(x).
\ees

\section { Realization on trajectories }
A trajectory in $(N|M)$-dimensional superspace is simply a
vector-valued function of time, $q^i(t)$, which satisfies the graded
Heisenberg algebra
together with its canonically conjugate momentum $p_j(t)$.
\be
[p_j(s), q^i(t)] = \dlt^i_j \dlt (s-t), \qquad
[q^i(s), q^j(t)] = [p_i(s), p_j(t)] = 0.
\label{traj}
\ee
By differentiating with respect to $t$, we obtain the useful relation
\be
[p_j(s), \dt q^i(t)] = - \dlt^i_j \dt\dlt (s-t).
\ee

The trajectories may formally be extended to super space-time, by defining
time components $q^0(t)$ and $p_0(t)$ as
\be
q^0(t) = t, \qquad
p_0(t) = -\dt q^i(t) p_i(t).
\ee
Clearly, $\deg q^0 = \deg p_0 = 0$.
The oscillators $q^\mu(t) = (q^0(t), q^i(t))$ and
$p_\nu(t) = (p_0(t), p_j(t))$ satisfy the following algebra.
\bes
[p_\nu(s), q^\mu(t)]
&=& (\dlt^\mu_\nu - \dlt^0_\nu \dt q^\mu(s)) \dlt (s-t), \nl
{[}p_\nu(s), \dt q^\mu(t)]
&=& -(\dlt^\mu_\nu - \dlt^0_\nu \dt q^\mu(s)) \dt\dlt (s-t), \nl
{[}q^\mu(s), q^\nu(t)] &=& [q^\mu(s), \dt q^\nu(t)] = 0 \nl
{[}p_\mu(s), p_\nu(t)]
&=& \bigg( \dlt^0_\mu p_\nu(s) + \dlt^0_\nu p_\mu(t) \bigg) \dt\dlt(s-t).
\label{osc}
\ees
Note that $p_0(t)$ satisfy $diff(1)$ (\ref{Vir}), and that
$\dt q^0(t) = 1$, $\ddt q^0(t) = \qmu(t) p_\mu(t) = 0$. 
Eq. (\ref{osc}) is formally the same as in the bosonic case. 
The only place where the super nature of these relations enter is that the
brackets are graded, e.g.
\be
[q^\mu(t), p_\nu(s)] = (-)^{\mu\nu} [p_\nu(s), q^\mu(t)].
\ee
It follows that
\be
[p_\mu(s), f(q(t))] = (\dmu f(q(t)) - \dlt^0_\mu \dt f(q(t))) \dlt(s-t).
\ee
We assume that super space-time is periodic in the temporal direction. 
This means that $\int dt\ \dt f(t) = 0$ for every (operator-valued) function;
in particular, if $\th$ is a fermionic coordinate, 
$\int dt\ \dt \th(t) = \int d\th 1 = 0$ can be interpreted as a property of
the
Berezin integral.
Moreover, every function can be expanded in a Fourier series.
The algebra (\ref{osc}) has a natural Fock module, which is obtained from
the universal envelopping algebra by introducing a vacuum which is
annihilated by all negative frequency modes.

\begin{theorem} \label{cLxithm} 
Let $L(t)$ satisfy $diff(1)$ and let $T^\nu_\mu(t)$ satisfy $map(1,
gl(\N|M))$.
Then the following expression provides a realization of $diff(\N|M)$.
\bes
\Lxi &=& \int dt \ \xmu(q(t)) p_\mu(t)
 + \xz(q(t)) L(t) + (-)^{(\xi+\mu)\nu+\mu} \dnu \xmu(q(t)) T^\nu_\mu(t) \nl
&=& \int dt \ \xi^i(q(t)) p_i(t) - \xz(q(t)) \dt q^i(t) p_i(t) \nl
&& + \xz(q(t)) L(t) + (-)^{(\xi+\mu)\nu+\mu} \dnu \xmu(q(t)) T^\nu_\mu(t),
\label{cLxi}
\ees
where $\xmu(q(t)) = \xmu(t, q^1(t), ..., q^{N+M}(t))$.
\end{theorem}
The proof is as in \cite{Lar97}, theorem 3.1, except for the extra signs.
\qed

Hence the following transformation law defines a $diff(\N|M)$ representation.
\bes
&&[\Lxi, \phi^{\si_1...\si_p}{}_{\tau_1...\tau_q}(t)]
= -\xz(q(t)) \dt\phi^{\si_1...\si_p}{}_{\tau_1...\tau_q}(t)
 - \la \dt\xz(q(t)) \phi^{\si_1...\si_p}{}_{\tau_1...\tau_q}(t) \nl
\bl + i\om\xz(q(t)) \phi^{\si_1...\si_p}{}_{\tau_1...\tau_q}(t)
 -\ka \div\xi(q(t)) \phi^{\si_1...\si_p}{}_{\tau_1...\tau_q}(t) \nl
\bl + \sum_{i=1}^p (-)^{(\si_1+..+\si_i)(\mu+\si_i)+\xi\mu}
 \dmu \xi^{\si_i}(q(t)) \phi^{\si_1..\mu..\si_p}{}_{\tau_1...\tau_q}(t) 
\label{prim} \\
\bl - \sum_{j=1}^q (-)^{(\si_1+..+\si_p+\tau_1+..+\tau_j+\mu)(\mu+\tau_j) +
\xi\tau_j}
 \d_{\tau_j}\xmu(q(t)) \phi^{\si_1...\si_p}{}_{\tau_1..\mu..\tau_q}(t).
\nonumber
\ees
We call $\phi^{\si_1...\si_p}{}_{\tau_1...\tau_q}(t)$ a 
{\em primary trajectory field } of type $\Prim(\la, \om; \ka, p, q)$.
The trajectory itself transforms as
\be
[\Lxi, q^\nu(t)] = \xi^\nu(q(t)) - \xz(q(t))\qnu(t).
\label{Lxiq}
\ee
Its time derivative is a primary trajectory field of type $\Prim(1,0;0,1,0)$.
From
\be
[\Lxi, \qnu(t)] = (-)^{\mu(\mu+\xi+\nu)} \dmu\xi^\nu(q(t))\qmu(t)
 - \dt\xz(q(t))\qnu(t) - \xz(q(t))\ddt q^\nu(t),
\label{Lxiqt}
\ee
it follows that $(-)^\nu \qnu(t)$ transforms as in (\ref{prim}). 

To understand the meaning of theorem \ref{cLxithm}, consider its restriction
to the spatial subalgebra generated by time-independent vector fields.
\be
\Lxi = \int dt \ \xi^i(\vq(t)) p_i(t)
 + (-)^{(\xi+i)j+i} \d_j \xi^i(\vq(t)) T^j_i(t)
\label{spat}
\ee
This is recognized as the action of infinitesimal diffeomorphisms on 
extended objects in superspace. There is no need here to limit ourselves to
one-dimensional objects; $t$ could very well have several components.
From the algebraic point of view, this action is highly reducible; 
in fact, for every value of $t$ we have an independent tensor density
(\ref{tensors}), and thus (\ref{spat}) describes a continuous direct sum
of tensor densities.

However, for one-dimensional extended objects two miracles occur.
First, we can extend the action to $(\N|M)$-dimensional super space-time by
means of (\ref{cLxi}). Now $t$ must be one-dimensional because the time
derivative appears, both in $p_0(t) = -\dt q^i(t)p_i(t)$ and in the 
right-hand side of (\ref{Vir}). This realization is no longer obviously
reducible, although it still {\em is} reducible in a more subtle manner. 
For a trajectory field of type $\Prim(1, \om; \ka, p, q)$ and smearing
functions $ f^{\tau_1...\tau_q}{}_{\si_1...\si_p}(x)$, define
\be
\phi(f) = \int dt\ f^{\tau_1...\tau_q}{}_{\si_1...\si_p}{}(q(t))
\phi^{\si_1...\si_p}{}_{\tau_1...\tau_q}(t).
\label{phif}
\ee
It can be shown that $\phi(f)$ transforms in the same fashion as
$\Phi(f)$ (\ref{Phif}), provided that $\Phi(x)$ has the weight $\ka-1$.
This phenomenon, which was called {\em correspondance} in \cite{Lar97},
implies that this type of trajectory field contains a tensor field
subrealization.

The second miracle is that one-dimensional
objects admit normal ordering, which gives rise to a superalgebra 
extension. We now proceed to calculate it. 
Split the delta function into positive and negative energy parts.
\be
\dlt^>(t) = \nnn \sum_{m>0} \e^{-imt}, \qquad
\dlt^<(t) = \nnn \sum_{m\leq0} \e^{-imt}.
\ee
\begin{lemma}{\label{deltalemma}} (\cite{Lar97}, Lemma 5.1)
\bes
&i.& \dlt^>(t) \dlt^<(-t) - \dlt^>(-t) \dlt^<(t)
= -\nnni\dt \dlt(t) \nl
&ii.& \dlt^>(t) \dt\dlt^<(-t) - \dt\dlt^>(-t) \dlt^<(t)
= {1\/ 4\pi i} (\ddt \dlt(t) + i\dt \dlt(t)) \nl
&iii.& \dt\dlt^>(t) \dt\dlt^<(-t) - \dt\dlt^>(-t) \dt\dlt^<(t)
= {1\/ 12\pi i} (\dddt \dlt(t) + \dt \dlt(t)) \nonumber
\ees
\end{lemma}
Introduce
\be
\txi^i(t) \equiv \txi^i(q(t),\dt q(t)) = \xi^i(q(t)) -\xz(q(t)) \dt q^i(t),
\label{txi}
\ee
and
\bes
{\chi^\gtrless_\xi}^i_j(t,s) 
&\equiv& [p^\gtrless_j(t), \txi^i(s)] \nl
&=& \d_j \txi^i(s) \dlt^\gtrless(t-s) 
 + (-)^{j\xi} \dlt^i_j \xz(s) \dt \dlt^\gtrless(t-s).
\ees
Moreover, set 
${\chi_\xi}^i_j(t,s) = {\chi^>_\xi}^i_j(t,s)+{\chi^<_\xi}^i_j(t,s)$.

\begin{lemma} \label{txilemma}
The expressions defined in (\ref{txi}) satisfy the following relations.
\bes
&&(-)^{\xi i+i} \d_i\txi^i 
= \div\xi - \dt \xz
\label{dtxi} \\
&&(-)^{(\xi+\eta+j)j} \d_j\dt\txi{}^i \d_i\teta^j 
= (-)^{(\xi+\eta+\nu)\nu } ( \dnu\dt\xmu\dmu\ynu 
  + \dnu\xz\qrho\drho\dt\ynu ) \nl
\bl  - \qrho\drho\dt\xmu\dmu\yz - \ddt\xz\dt\yz - \dt\xz\qrho\drho\dt\yz
  +\qrho\drho\dt\xz\dt\yz \nl
\bl  + {d\/dt}(\dt\xz\dt\yz - (-)^{(\xi+\eta+\nu)\nu} \dnu\xz\dt\ynu).
\label{dydx}
\ees
\end{lemma}
{\em Proof: }
We use that $\txi^0 \equiv 0$.
Eq. (\ref{dtxi}) thus equals
\be
(-)^{\xi\mu+\mu} \dmu\txi^\mu = (-)^{\xi\mu+\mu} (\dmu\xmu - \dmu\xz\qmu),
\ee
whereas (\ref{dydx}) becomes
\bes
&&(-)^{(\xi+\eta+\nu)\nu} \dnu\dt\txi{}^\mu \dmu\teta^\nu \nl
&=&(-)^{(\xi+\eta+\nu)\nu}(\dnu\dt\xmu-\dnu\xz\ddt q^\mu - \dnu\dt\xz\qmu)
  (\dmu\ynu - \dmu\yz\qnu) \nl
&=& (-)^{(\xi+\eta+\nu)\nu} (\dnu\dt\xmu\dmu\ynu 
 - \dnu\xz(\ddt\ynu-\qrho\drho\dt\ynu) - \dnu\dt\xz\dt\ynu) \nl
 &&- \qrho\drho\dt\xmu\dmu\yz + \dt\xz(\ddt\yz-\qrho\drho\dt\yz)
  + \qrho\drho\dt\xz\dt\yz.
\qed
\ees

Normal ordering amounts to the replacement
\be
\txi^i(t) p_i(t) \rightarrow \no{ \txi^i(t) p_i^<(t) }
 = \txi^i(t) p_i^<(t) + (-)^{\xi i+i} p_i^>(t) \txi^i(t).
\ee
Moreover, it also affects the generators of $diff(1)$ and
$map(1, gl(\N|M))$, replacing these algebras by their central extensions.

\begin{theorem}{\label{xLxithm}} 
Let $L(t)$ satisfy the Virasoro algebra $Vir$ (\ref{Vir}) with central
charge $c$ and let $T^\mu_\nu(t)$ satisfy the Kac-Moody super-algebra 
$\km{gl(\N|M)}$ (\ref{kmgln}) with central charges $k_1$ and $k_2$.
The generators
\bes
\Lxi &=& \int dt \ \no{ \xmu(q(t)) p_\mu(t) } 
 + \xz(q(t)) L(t) + (-)^{(\xi+\mu)\nu+\mu} \dnu \xmu(q(t)) T^\nu_\mu(t) \nl
&\equiv& \int dt \ \txi^i(q(t),\dt q(t)) p_i^<(t) 
 + (-)^{\xi i+i} p_i^>(t) \txi^i(q(t),\dt q(t)) \nl
&& + \xz(q(t)) L(t) + (-)^{(\xi+\mu)\nu+\mu} \dnu \xmu(q(t)) T^\nu_\mu(t)
\label{Lxi}
\ees
satisfy the superalgebra 
\be
[\Lxi,\Leta]=\L_\bxy + \ext(\xi,\eta).
\label{Lxy}
\ee
The extension is
\bes
&&\ext(\xi,\eta) = \nnni \int dt\ \bigg\{ 
  (1+k_1) (-)^{(\xi+\eta+\nu)\nu} \dnu\dt\xmu(q(t)) \dmu\ynu(q(t)) \nl
 \bl + k_2 \div\dt\xi(q(t)) \div\eta(q(t)) \nl
 \bl +(-)^{(\xi+\eta+\nu)\nu} \dnu\xz(q(t)) \qrho(t)\drho\dt\ynu(q(t))
 -\qrho(t)\drho\dt\xmu(q(t)) \dmu\yz(q(t))  \nl
 \bl- \dt\xz(q(t)) \qrho(t)\drho\dt\yz(q(t)) 
 + \qrho(t)\drho\dt\xz(q(t)) \dt\yz(q(t)) 
\label{ext} \\
 \bl  + \half \div \dt\xi(q(t)) \dt\yz(q(t)) 
  - \half \dt\xz(q(t)) \div\dt\eta(q(t)) \nl
 \bl  - (2-{c+2(N-M)\/12}) \ddt\xz(q(t)) \dt\yz(q(t)) \nl
 \bl - {c+2(N-M)\/12}\dt\xz(q(t)) \yz(q(t)) \nl
 \bl  + {i\/ 2} ( \div \xi(q(t)) \dt\yz(q(t))
  - \dt\xz(q(t)) \div\eta(q(t)) ) \bigg\} , \nonumber
\ees
where $\dt f(q(t)) = \qrho(t)\drho f(q(t))$, $\div\xi$ was defined in 
(\ref{div}) and $q^\mu(t)$ transforms as in (\ref{Lxiq}).
\end{theorem}

{\em Proof:\ }
We begin by considering $\Lzxi = \int dt\ \no{ \txi^i(t) p_i(t) }$.
\bes
[\Lzxi, \Lzeta] &=& \iint ds dt\
 [\txi^i(s) p_i^<(s) + (-)^{\xi i+i} p_i^>(s) \txi^i(s), \nl
 \bl\teta^j(t) p_j^<(t) + (-)^{\eta j+j} p_j^>(t) \teta^j(t)] \nl
&=& \iint ds dt\ 
  \txi^i(s) \chi^{<j}_{\eta i}(s,t)p^<_j(t) \nl
 \bl   +(-)^{\xi(\eta+j)+ij} \teta^j(t) 
   (-(-)^{(\xi+i)j}\chi^{< i}_{\xi j}(t,s))p^<_i(s) \nl
&& +(-)^{\eta j+j} \bigg\{ 
  (-)^{ij} \txi^i(s) p^>_j(t) \chi^{<j}_{\eta i}(s,t) \nl
  \bl  +(-)^{i\eta} (-(-)^{j(\xi+i)} \chi^{> i}_{\xi j}(t,s))\teta^j(t)
  p^<_i(s) \bigg\} \nl
&& + (-)^{\xi i+i} \bigg\{ 
  (-)^{(\xi+i)\eta} \chi^{> j}_{\eta i}(s,t)p^<_j(t)\txi^i(s)\nl
  \bl   + (-)^{(\xi+i)(\eta+j)} p^>_i(s) \teta^j(t) 
    (-(-)^{(\xi+i)j} \chi^{< i}_{\xi j}(t,s) ) \bigg\} \nl
&& +(-)^{\xi i+i+\eta j+j} \bigg\{
    p^>_i(s) (-(-)^{j(\xi+i)} \chi^{> i}_{\xi j}(t,s)) \teta^j(t) \nl
  \bl   + (-)^{\xi j+(\xi+i)(\eta+j)} p^>_j(t)\chi^{> j}_{\eta i}(s,t)
  \txi^i(s) \bigg\}.
\ees
Of these eight terms, the third can be rewritten as
\be
(-)^{(\eta+j+i)j} \bigg\{ (-)^{(\xi+i)j} p^>_j(t) \txi^i(s) \chi^{<j}_{\eta
i}(s,t)
 -(-)^{(\xi+i)j}  \chi^{> i}_{\xi j}(t,s) \chi^{<j}_{\eta i}(s,t) \bigg\}
\ee
and the fifth as
\be
(-)^{(\xi+i)(\eta+i)}\bigg\{ (-)^{(\xi+i)j} \chi^{> j}_{\eta i}(s,t)
\txi^i(s)p^<_j(t)
 + \chi^{> j}_{\eta i}(s,t)\chi^{< i}_{\xi j}(t,s) \bigg\}
\ee
Hence 
\bes
&&[\Lzxi, \Lzeta] = \iint ds dt\
 \txi^i(s) \chi^{<j}_{\eta i}(s,t)p^<_j(t)
  - (-)^\xy  \teta^j(t) \chi^{< i}_{\xi j}(t,s)p^<_i(s) \nl
\bl+(-)^{(\xi+\eta+j)j} p^>_j(t) \txi^i(s) \chi^{<j}_{\eta i}(s,t) 
  - (-)^\xy \teta^j(t) \chi^{> i}_{\xi j}(t,s) p^<_i(s) \nl
\bl+  \txi^i(s)\chi^{> j}_{\eta i}(s,t)p^<_j(t) 
  - (-)^{(\xi+i)(\eta+i)} p^>_i(s) \teta^j(t) \chi^{< i}_{\xi j}(t,s) \\
\bl-(-)^{(\xi+i)(i+\eta)} p^>_i(s) \teta^j(t) \chi^{> i}_{\xi j}(t,s)
  + (-)^{(\xi+\eta+j)j} p^>_j(t) \txi^i(s) \chi^{> j}_{\eta i}(s,t)) \nl
\bl- (-)^{(\xi+\eta+j)j}  \chi^{> i}_{\xi j}(t,s) \chi^{<j}_{\eta i}(s,t)
  + (-)^{\xi\eta+(\xi+\eta+i)i} \chi^{> j}_{\eta i}(s,t)\chi^{< i}_{\xi
  j}(t,s) \bigg\}.
\nonumber
\ees
The regular piece is
\be
\iint ds dt\ \txi^i(s) {\chi_\eta}^j_i(s,t) p^<_j(t) 
 + (-)^{(\xi+\eta+j)j} p^>_j(t) \txi^i(s){\chi_\eta}^j_i(s,t) - \xcy.
\ee
We focus on the first term.
\bes
&&\iint ds dt\ \txi^i(s) {\chi_\eta}^j_i(s,t) p^<_j(t) - \xcy \nl
&=& \iint ds dt\ \txi^\mu(s) ( \dmu\teta^j(t) \dlt(s-t)
 +(-)^{\mu\eta} \yz(t) \dlt^j_\mu \dt\dlt(s-t) ) p^<_j(t) - \xcy \nl
&=& \int \bigg\{ (\widetilde{ \xmu\dmu\eta})^j 
 - \xz(\dt\eta^j - \dt\yz\dt q^j) - (-)^{j\eta} \dt\txi{}^j \yz 
  \bigg\} p^<_j(t) - \xcy,
\ees
which equals $\Lz_\bxy$. This 
could have been anticipated from theorem \ref{cLxithm}.
We here suppressed the integration variable in the single integral, because
no confusion is possible.
The extension $\ext_0(\xi,\eta)$ becomes
\bes
&& \iint ds dt \
 -(-)^{(\xi+\eta+j)j} \chi^{>i}_{\xi j}(t,s) \chi^{<j}_{\eta i}(s,t)
 + (-)^{\xy+(\xi+\eta+i)i} \chi^{>j}_{\eta i}(s,t) \chi^{<i}_{\xi j}(t,s) \nl
&=& - \iint ds dt \
 (-)^{(\xi+\eta+j)j} (\d_j \txi^i(s) \dlt^>(t-s)
 + (-)^{j\xi} \dlt^i_j \xz(s) \dt \dlt^>(t-s))\times \nl
 && \times (\d_i \teta^j(t) \dlt^<(s-t)
  + (-)^{i\eta} \dlt^j_i \yz(t) \dt \dlt^<(s-t) ) - \xcy \nl
&=& - \iint ds dt \
 (-)^{(\xi+\eta+j)j} \d_j \txi^i(s) \d_i \teta^j(t) \dlt^>(t-s) \dlt^<(s-t)
 \nl
 &&+ (-)^{(\eta+j)j} \xz(s) \d_j \teta^j(t) \dt\dlt^>(t-s) \dlt^<(s-t) \nl
 &&+ (-)^{(\xi+i)i} \d_i \txi^i(s) \yz(t) \dlt^>(t-s) \dt\dlt^<(s-t) \nl
 &&+ (-)^i \dlt^i_i \xz(s) \yz(t)\dt\dlt^>(t-s) \dt\dlt^<(s-t) - \xcy \nl
&=& \nnni \iint ds dt \
  (-)^{(\xi+\eta+j)j} \d_j \txi^i(s) \d_i \teta^j(t) \dt\dlt(t-s) \nl
 &&+ \half (-)^{(\eta+j)j} \xz(s) \d_j \teta^j(t) (\ddt\dlt(t-s) - i\dt
 \dlt(t-s)) \nl
 &&- \half (-)^{(\xi+i)i} \d_i \txi^i(s) \yz(t) (\ddt\dlt(t-s) + i\dt
 \dlt(t-s)) \nl
 &&- {N-M\/6} \xz(s) \yz(t) (\dddt\dlt(t-s) + \dt \dlt(t-s)) \nl
&=& \nnni \int 
 (-)^{(\xi+\eta+j)j} \d_j \dt\txi{}^i \d_i \teta^j
 - \half (-)^{(\eta+j)j} \dt\xz \d_j \dt\teta{}^j 
 + \half (-)^{(\xi+i)i} \d_i \dt\txi{}^i \dt\yz \nl
 &&- {N-M\/6} ( -\ddt\xz \dt\yz + \dt\xz\yz ) \nl
 &&+ {i\/ 2} ( -(-)^{(\eta+j)j} \dt\xz \d_j \teta^j
  +(-)^{(\xi+i)i} \d_i \txi^i \dt\yz ),
\label{ext0}
\ees
where we used Lemma \ref{deltalemma} and the fact that 
$(-)^i \dlt^i_i = N-M$.
Now consider the full algebra.
\bes
&&[\Lxi, \Leta] \equiv [\Lxi, \Leta] + \ext(\xi,\eta) \nl
&=& \L_\bxy + \ext_0(\xi,\eta)
 + {c\/24\pi i} \iint ds dt\ \xz(s)\yz(t) (\dddt\dlt(s-t)+ \dt\dlt(s-t)) \nl
 &&- \nnni \iint ds dt\ \dsi\xmu(s) \d_\tau\ynu(t)
  (-)^{(\xi+\mu)\si+\mu+(\eta+\nu)\tau+\nu} \times \nl
 \bl  (-)^{(\si+\mu)(\eta+\nu+\tau)} 
  (k_1 (-)^\si \dlt^\si_\nu \dlt^\tau_\mu
   + k_2 (-)^{\si+\tau} \dlt^\si_\mu \dlt^\tau_\nu )\dt\dlt(s-t).
\ees
Thus,
\bes
&&\ext(\xi,\eta)
= \ext_0(\xi,\eta) + \nnni \int {c\/12}(\ddt\xz\dt\yz - \dt\xz\yz) \nl
\bl  + k_1 (-)^{(\xi+\eta+\nu)\nu} \dnu\dt\xmu\dmu\ynu 
 + k_2 (-)^{\xi\mu+\mu+\eta\nu+\nu} \dmu\dt\xmu\dnu\ynu.
\ees
The result now follows by means of lemma \ref{txilemma}.
As a consistency check we note that the extension satisfies 
$\ext(\eta,\xi) = -(-)^\xy \ext(\xi,\eta)$.
\qed

The superalgebra described in this theorem is not a Lie superalgebra, because
the right-hand side is not linear in $q^\mu(t)$. Rather, it is a graded
associative algebra, and the bracket must be interpreted as the graded
commutator.
However, it is easy to rewrite the extension in linear form, by introducing
a sufficient number of new generators. We then obtain an abstract Lie 
superalgebra extension of $diff(\N|M)$.

Let $h = h_\mun(x) dx^{\mu_1}\circ..\circ dx^{\mu_n}$ be a graded 
symmetric $n$-tensor and let $g = g_\mu(x) dx^\mu$ be a one-tensor. 
Define the operators
\bes
S_n(h) &=& -\nnni\int dt\ \dt q^{\mu_1}(t) .. \dt q^{\mu_n}(t) h_\mun(q(t))
\nl
R_n(g, h) &=& -\nnni\int dt\
\ddt q^\mu(t) \dt q^{\nu_1}(t) .. \dt q^{\nu_n}(t) g_\mu(q(t)) h_\nun(q(t)).
\label{SRq}
\ees
We now proceed somewhat differently from \cite{Lar97}, in order to keep 
better track of the minus signs. Define kernels $S_n^\nun(x)$ and
$R_n^{\mu|\nun}(x)$ by
\bes
S_n(h) &=& -\int \dx\ S_n^\nun(x) h_\nun(x) \nl
R_n(g, h) &=& -\int \dx\ R_n^{\mu|\nun}(x) g_\mu(x) h_\nun(x).
\label{SR}
\ees
Both fields are graded symmetric in $\nun$.
To show that these definitions are consistent, i.e. that  (\ref{SRq}) and
(\ref{SR})
transform identically, we integrate by parts and throw away the boundary
terms.
Thus, we assume that the relations
\be
\int dt\ \dt f(t) = \int \dx\ \dmu F(x) = 0
\ee
hold for all functions $f(t)$ and $F(x)$. 
The kernels must satisfy the following relations.
\bes
&&[\Lxi, S_n^\nun(x)] = 
 -\xmu(x)\dmu S_n^\nun(x) - \div\xi(x) S_n^\nun(x) \nl
 \bl  + \sum_{j=1}^n (-)^{(\mu+\nu_j)(\sumnu{j-1}+\mu)+\xi\mu}
   \dmu\xi^{\nu_j}(x) S_n^{\nu_1..\mu..\nu_n}(x) \nl
 \bl- (n-1) (-)^{\xi\mu+\mu} \dmu\xz(x) S_{n+1}^{\mu\nun}(x), \nl
&&(-)^\nu \dnu S_1^\nu(x) = 0, \nl
&&S_{n+1}^{0\nun}(x) = S_n^\nun(x),
\label{Sn}
\ees
and
\bes
&&[\Lxi, R_n^{\si|\nun}(x)] = 
 -\xmu(x)\dmu R_n^{\si|\nun}(x) - \div\xi(x) R_n^{\si|\nun}(x) \nl
 \bl  + (-)^{\mu(\xi+\si+\mu)} \dmu\xi^\si(x) R_n^{\mu|\nun}(x) \nl
 \bl+ \sum_{j=1}^n (-)^{(\mu+\nu_j)(\si+\sumnu{j-1}+\mu)+\xi\mu} 
  \dmu\xi^{\nu_j}(x) R_n^{\si|\nu_1..\mu..\nu_n}(x) \nl
 \bl  - (n+1) (-)^{\mu(\xi+\si+\mu)} \dmu\xz(x) R_{n+1}^{\si|\mu\nun}(x) \nl
 \bl - (-)^{\xi\mu+\mu} \dmu\xz(x) R_{n+1}^{\mu|\si\nun}(x) \nl
 \bl  + (-)^{(\mu+\rho)(\xi+\si+\rho)+\mu}\dmu\drho\xi^\si(x)
 S_{n+2}^{\mu\rho\nun}(x) \nl
 \bl - (-)^{(\mu+\rho)(\xi+\rho)+\mu} \dmu\drho\xz(x)
 S_{n+3}^{\mu\rho\si\nun}(x), \nl
&&(-)^\mu\dmu S_{n+1}^{\mu\nun}(x) = 
  \sum_{j=1}^n (-)^{\nu_j(\sumnu{j-1})} 
   R_{n-1}^{\nu_j|\nu_1..\check\nu_j..\nu_n}(x), \nl
&&R_{n+1}^{\si|0\mun}(x) = R_n^{\si|\mun}(x), \nl
&&R_n^{0|\nun}(x) = 0,
\label{Rn}
\ees
where the check mark $\check\nu_j$ denotes omission.
The subsidiary conditions follow from $\int dt\ \dt f(q(t)) = 0$, $\dt q^0(t)
= 1$,
$\int dt\ {d\/dt}(\dt q^{\nu_1}(t)..\dt q^{\nu_n}(t)h_\nun(q(t))) = 0$, 
$\dt q^0(t) = 1$, and $\ddt q^0(t) = 0$, respectively.
The extension can now be rewritten as
\bes
&&\ext(\xi,\eta) = \nnni \int dt\ \bigg\{ 
 (1+k_1) (-)^{(\xi+\eta+\nu)\nu} \qrho\drho\dnu\xmu\dmu\ynu \nl
 \bl+k_2 \qrho\drho\div\xi\div\eta 
  +(-)^{(\xi+\eta+\nu)\nu}(-)^{\rho(\nu+\xi)+\si(\nu+\xi+\rho)}
  \qrho\qsi\dnu\xz\drho\dsi\ynu \nl
 \bl - (-)^{\si\rho} \qrho\qsi\drho\dsi\xmu\dmu\yz 
 - (-)^{\rho\xi+\tau(\si+\xi+\rho)} \qrho\qsi\qtau \dsi\xz\drho\dtau\yz \nl
 \bl + (-)^{\si\rho+\tau(\rho+\si+\xi)} \qrho\qsi\qtau\drho\dsi\xz\dtau\yz\nl
 \bl + \half (-)^{\si(\rho+\xi)} \qrho\qsi \drho\div\xi\dsi\yz 
 - \half (-)^{\si(\rho+\xi)} \qrho\qsi \drho\xz\dsi\div\eta\nl
 \bl - (2-{c+2(N-M)\/12}) ( (-)^{\si(\rho+\xi)} \ddt
 q^\rho\qsi\drho\xz\dsi\yz \nl
  \bl+ (-)^{\rho\si+\tau(\rho+\si+\xi)} \qrho\qsi\qtau \drho\dsi\xz\dtau\yz ) 
  - {c+2(N-M)\/12}\qrho\drho\xz\yz \nl
 \bl + {i\/ 2} ( (-)^{\xi\rho}\qrho \div\xi\drho\yz
  - \qrho\drho\xz\div\eta ) \bigg\} \nl
&&= \int \dx\ \bigg\{ 
 (1+k_1) (-)^{(\xi+\eta+\nu)\nu} S_1^\rho(x) \drho\dnu\xmu(x) \dmu\ynu(x) \nl
 \bl + k_2  S_1^\rho(x) \drho\div\xi(x) \div\eta(x) \nl
 \bl +(-)^{(\xi+\eta+\nu)\nu+\rho(\nu+\xi)+\si(\nu+\xi+\rho)}
  S_2^{\rho\si}(x) \dnu\xz(x) \drho\dsi\ynu(x) \nl
 \bl - (-)^{\si\rho} S_2^{\rho\si}(x) \drho\dsi\xmu(x) \dmu\yz(x) \nl
 \bl- (-)^{\rho\xi+\tau(\si+\xi+\rho)} S_3^{\rho\si\tau}(x) \dsi\xz(x)
 \drho\dtau\yz(x) \nl
 \bl + (-)^{\si\rho+\tau(\rho+\si+\xi)} S_3^{\rho\si\tau}(x) \drho\dsi\xz(x)
 \dtau\yz(x) \nl
 \bl + \half (-)^{\si(\rho+\xi)} S_2^{\rho\si}(x) 
   ( \drho\div\xi(x) \dsi\yz(x) - \drho\xz(x) \dsi\div\eta(x) ) \nl
 \bl - (2-{c+2(N-M)\/12}) ( (-)^{\si(\rho+\xi)} R_1^{\rho|\si}(x) \drho\xz(x)
 \dsi\yz(x) \nl
  \bl + (-)^{\rho\si+\tau(\rho+\si+\xi)} S_3^{\rho\si\tau}(x) \drho\dsi\xz(x)
  \dtau\yz(x) )\nl
 \bl - {c+2(N-M)\/12} S_1^\rho(x) \drho\xz(x) \yz(x) \nl
 \bl + {i\/ 2} ( (-)^{\xi\rho} S_1^\rho(x) \div\xi(x)\drho\yz(x)
  - S_1^\rho(x) \drho\xz(x) \div\eta(x) ) \bigg\} 
\label{xext}
\ees
Eqs. (\ref{Lxy}), (\ref{Sn}), (\ref{Rn}) and (\ref{xext}), together with
the conditions 
\bes
&&[S_m^\mum(x), S_n^\nun(y)] =
[S_m^\mum(x), R_n^{\si|\nun}(y)] \nl
&&= [R_m^{\si|\mum}(x), R_n^{\tau|\nun}(y)] = 0,
\ees
define an abstract Lie superalgebra, and the expressions in (\ref{Lxi})
and (\ref{SRq}) provide a realization of it. 
Further, if we represent (\ref{osc}) on the natural graded Fock module,
and pick arbitrary lowest-weight modules for $Vir$ and 
$\km{gl(\N|M)}$, we obtain a lowest-energy module for the non-centrally
extended super-diffeomorphism algebra.

\section {Subalgebras}
A projective representation of $diff(\N|M)$ yields by restriction a
projective
representation of its subalgebras, i.e. superalgebras whose generators admit
a realization as first-order differential operators on super space-time.
Such algebras are described in \cite{BL81}\cite{GLS97}. The bosonic case 
is classical, and can be found e.g. in \cite{Fuks86}.

\subsection{ Temporal subalgebra $diff(1)$ }
The {\em temporal subalgebra} is generated by space-independent vector fields
$\xi = \xi^0(t)\d_0$. Eq. (\ref{Lxi}) becomes
\bes
\Lxi &=& \int dt\ \xz(t) (-\no{\dt q^i(t)p_i(t)} + L(t)) + \dt\xz(t) T^0_0(t)
\nl
&=& \int dt\ \xz(t) L'(t)
\ees
where
\be
L'(t) \equiv -\no{\dt q^i(t)p_i(t)} + L(t) - \dt T^0_0(t)
\ee
generates a Virasoro algebra. The extension (\ref{ext}) is
\be
\ext(\xi,\eta) = \nnni \int (k_1+k_2+{c+2(N-M)\/12}) \ddt\xz\dt\yz
 - {c+2(N-M)\/12} \dt\xz\yz .
\ee
Hence the temporal subalgebra is a Virasoro algebra with central charge 
$c_{Temp} = c + 2(N-M) + 12(k_1+k_2)$.

\subsection{ Spatial subalgebra $diff(N|M)$ }
The {\em spatial subalgebra} is generated by time-independent vector fields 
$\xi = \xi^i(\vx) \d_i$. The arrow denotes a vector with spatial components
only;
$\vx^0 = 0$. Using that $\xi^0 = \d_0\xi^i = 0$, we find
\be
\Lxi = \int dt\ \no{ \xi^i(\vq(t)) p_i(t) } 
  + (-)^{(\xi+i)j+i} \d_j \xi^i(\vq(t)) T^j_i(t).
\ee
The extension becomes
\bes
\ext(\xi,\eta) &=& \nnni\int dt\
 (1+k_1)(-)^{(\xi+\eta+j)j} \d_j\dt\xi^i(\vq(t)) \d_i\eta^j(\vq(t)) \nl
 &&+ k_2(-)^{\xi i+i+\eta j+j}  \d_i\dt\xi^i(\vq(t)) \d_j \eta^j(\vq(t)).
\ees

\subsection{ Special superdiffeomorphism algebra $sdiff(\N|M)$ }
It can be shown that the divergence (\ref{div}) of a vector field satisfies
\be
\div([\xi,\eta]) = \xmu\dmu\div\eta - (-)^\xy \ynu\dnu\div\xi.
\ee
The {\em special} (or {\em divergence-free}) algebra $sdiff(\N|M)$ is
generated by vector fields with vanishing divergence. 
In theorems \ref{cLxithm} and \ref{xLxithm}, the matrices 
$T^\mu_\nu(t) \in sl(\N|M)$, and hence
$k_1 = (N-M+1)k_2$ in (\ref{kmgln}).
The extension is
\bes
&&\ext(\xi,\eta) = \nnni \int dt\ \bigg\{ 
  (1+k_1) (-)^{(\xi+\eta+\nu)\nu} \dnu\dt\xmu(q(t)) \dmu\ynu(q(t)) \nl
 \bl +(-)^{(\xi+\eta+\nu)\nu} \dnu\xz(q(t)) \qrho(t)\drho\dt\ynu(q(t))
 -\qrho(t)\drho\dt\xmu(q(t)) \dmu\yz(q(t))  \nl
 \bl- \dt\xz(q(t)) \qrho(t)\drho\dt\yz(q(t)) 
 + \qrho(t)\drho\dt\xz(q(t)) \dt\yz(q(t)) \nl
 \bl  - (2-{c+2(N-M)\/12}) \ddt\xz(q(t)) \dt\yz(q(t)) \nl
 \bl - {c+2(N-M)\/12}\dt\xz(q(t)) \yz(q(t)) \bigg\}.
\label{sext}
\ees

\subsection{ Hamiltonian algebras $H(\N|M)$ and $H(N|M)$}
The Hamiltionian algebra $H(\N|M)$ preserves the constant graded
skew-symmetric
matrix $\om_{\si\tau}$, satisfying
\be
\om_{\tau\si} = -(-)^{\si\tau} \om_{\si\tau}
\ee
Actually, it is sufficient if $\om_{\si\tau}(x)$ is a closed two-form, but we
do not need
this generalization here, because it is always possible to choose Darboux
coordinates locally. However, it would be necessary for global
considerations.
Define the inverse matrix $\om^{\mu\nu}$ by
\be
\om^{\mu\rho}\om_{\rho\nu} = (-)^\mu\dlt^\mu_\nu, \qquad
(-)^\rho \om_{\nu\rho} \om^{\rho\mu} = \dlt^\mu_\nu
\ee
In general, a two-form transforms as
\bes
[\Lxi, \om_{\si\tau}(x)] &=& - \xmu\dmu\om_{\si\tau}(x)
 - (-)^{\xi\si+\si+\mu} \dsi\xmu\om_{\mu\tau}(x) \nl
 &&- (-)^{\xi\tau+\tau+\mu+\si(\mu+\tau)}\dtau\xmu\om_{\si\mu}(x).
\label{2form}
\ees
The matrix $\om_{\si\tau}$ can be regarded as a constant two-form, provided
$\xi$ is a Hamiltonian vector field of the form
\be
\xi = H_f \equiv (-)^{f\mu+\nu} \dmu f(x)\om^{\mu\nu}\dnu,
\label{Hf}
\ee
for $f$ an arbitary function. Such vector fields generate the Hamiltonian
algebra $H(\N|M) \subset diff(\N|M)$. 
It is easy to verify that (\ref{Hf}) and
(\ref{2form}) imply that $[\Lxi, \om_{\si\tau}(x)] = 0$, and that
\bes
[H_f, H_g] &=& H_{\{f,g\}_\PB}, \nl
\{f, g\}_\PB &=& (-)^{f\mu+\nu} \dmu f \om^{\mu\nu} \dnu g.
\label\PB
\ees
$\{\cdot,\cdot\}_\PB$ is called the Poisson bracket. It satisfies the
axioms of a Lie superalgebra, acting as a derivation of the associative
product.
\bes
&&\{g,f\}_\PB = (-)^{fg} \{f, g\}_\PB \nl
&&\{f,gh\}_\PB = \{f,g\}_\PB\ h + (-)^{fg} g \{f,h\}_\PB \nl
&&(-)^{fh} \{f, \{g, h\}_\PB\}_\PB
 +(-)^{fg} \{g, \{h, f\}_\PB\}_\PB \nl
 &&\qquad + (-)^{gh} \{h, \{f, g\}_\PB\}_\PB = 0.
\ees
Conventionally, one sets $\om^{\mu\nu}=0$ if $\deg\ \mu+\deg\ \nu = 1$. 
However, it is only necessary to demand that $\om^{\mu\nu}$ be a Grassmann 
(anti-commuting) number in this case. It appears that by choosing
$\om^{\mu\nu}$
purely Grassmann, the Leitesian algebra \cite{Lei77}, \cite{GLS97}, i.e. the
odd analogue of the Hamiltonian algebra, is obtained.

Inserting (\ref{Hf}) into (\ref{Lxi}) yields the following realization
\bes
\L(H_f) &=& \int dt \ 
 (-)^{f\mu+\nu} \no{ \dmu f(q(t)) \om^{\mu\nu} p_\nu(t) } 
\label{LHf} \\
 &&+ (-)^{f\mu} \dmu f(q(t)) \om^{\mu0} L(t) 
 + (-)^{f(\mu+\rho)+\nu\rho} \drho\dmu f(q(t)) \om^{\mu\nu} T^\rho_\nu(t).
\nonumber
\ees
We have $\deg\ H_f = \deg\ f$ and $H_f^\mu = (-)^{f\si+\mu} \dsi
f\om^{\si\mu}$.
Moreover, 
\be
\div(H_f) = (-)^{f\nu+\nu+f\mu+\nu} \dnu\dmu f \om^{\mu\nu} = 0.
\ee
The generators in (\ref{LHf}) satisfy an extended Hamiltonian algebra,
\be
[\L(H_f), \L(H_g)] = \L(H_{\{f,g\}_\PB}) + \ext(f,g),
\ee
where the extension is obtained by specialization of (\ref{ext}).
\bes
&&\ext(f,g) \equiv \ext(H_f, H_g) \nl
&&= {(-)^{f\si+g\tau}\/2\pi i} \int dt\ 
 (1+k_1) (-)^{(f+g)\nu+\mu} \dnu\dsi\dt f \om^{\si\mu} \dmu\dtau g
 \om^{\tau\nu} \nl
\bl+ (-)^{(f+g)\nu} \dnu\dsi f \om^{\si0}\qrho\drho\dtau\dt g\om^{\tau\nu} 
- (-)^\mu \qrho\drho\dsi\dt f \om^{\si\mu}\dmu\dtau g \om^{\tau0} \nl
\bl- \dsi\dt f\om^{\si0} \qrho\drho\dtau\dt g\om^{\tau0} 
+ \qrho\drho\dsi\dt f\om^{\si0} \dtau\dt g \om^{\tau0} \nl
\bl- (2-{c+2(N-M)\/12}) \dsi\ddt f \om^{\si0} \dtau\dt g\om^{\tau0} \nl
\bl- {c+2(N-M)\/12} \dsi\dt f\om^{\si0} \dtau g\om^{\tau0}.
\ees
where $f = f(q(t))$ and $g = g(q(t))$.

The dimensions in the Hamiltonian algebra appear in pairs: a coordinate and
its conjugate momentum. Since time is a distinguished dimension, it is
natural to
consider matrices satisfying $\om^{0\nu} = 0$, and time-independent
functions.
This leads to considerable simplifications.
\bes
\{f, g\}_\PB &=& (-)^{fi+j} \d_i f(\vx) \om^{ij} \d_j g(\vx), 
\label{pb0} \\
H_f &=& (-)^{fi+j} \d_i f(\vx)\om^{ij}\d_j, 
\label{Hf0} \\
\L(H_f) &=& \int dt \ (-)^{fi+j} \no{ \d_i f(\vq(t))\om^{ij} p_j(t) } \nl
  &&+ (-)^{f(i+k)+jk} \d_k\d_i f(\vq(t)) \om^{ij} T^k_j(t), \\
\ext(f,g) &=& (1+k_1) {(-)^{fk+gl+(f+g)j+i}\/2\pi i} \times \nl
&&\times \int dt\ \d_j\d_k\dt f(\vq(t)) \om^{ki} \d_i\d_\ell g(\vq(t))
\om^{\ell j}.
\ees

\subsection{ Contact algebra $K(\N|M)$ }

Denote the Euler operator $E = x^i\d_i$ and $\Delta = 2-E$. 
Clearly, $\deg \Delta = \deg E = \deg \d_0 = 0$.
The contact algebra $K(\N|M)$ is
\be
[K_f, K_g] = K_{\{f,g\}_\KB}, 
\ee
where
\be
\{f, g\}_\KB = \Delta(f(x)) \d_0g(x) - \d_0f(x) \Delta(g(x)) - \{f, g\}_\PB,
\label{KB}
\ee
and the Poisson bracket is given by (\ref{pb0}).
$\{\cdot,\cdot\}_\KB$ is called the contact bracket. It satisfies the
axioms of a Lie superalgebra, but it is not a derivation of the associative 
product, due to an extra term.
\bes
&&\{g,f\}_\KB = (-)^{fg} \{f, g\}_\KB \nl
&&\{f,gh\}_\KB = \{f,g\}_\KB\ h + (-)^{fg} g \{f,h\}_\KB + 2\d_0 f\ gh \nl
&&(-)^{fh} \{f, \{g, h\}_\KB\}_\KB +
 (-)^{fg} \{g, \{h, f\}_\KB\}_\KB \nl
 &&\qquad + (-)^{gh} \{h, \{f, g\}_\KB\}_\KB = 0.
\ees
To verify that (\ref{KB}) defines a Lie algebra, the following formulas are
useful.
\bes
\d_0 \Delta &=& \Delta \d_0 \nl
\Delta \{f, g\}_\PB &=& \{\Delta f, g\}_\PB + \{f, \Delta g\}_\PB \nl
\Delta(fg) &=& \Delta(f)g + f\Delta(g) - 2fg
\ees
$K(\N|M)$ is realized by contact vector fields
\be
K_f = \Delta(f(x)) \d_0 - H_f + \d_0f(x) E
\ee
which is verified by direct computation.
The components are
\bes
K_f^0(x) &=& \Delta(f(x)) \nl
K_f^i(x) &=& -H^i_f(x) + \d_0f(x)x^i, 
\label{Kf}
\ees
where 
\be
H^i_f(x) = (-)^{fk+i} \d_k f(x)\om^{ki}.
\ee
The realization on trajectories is now obtained by substituting (\ref{Kf})
into (\ref{Lxi}), and making the replacements
$\d_0 \mapsto p_0(t) = - \dt q^i(t) p_i(t)$, $E \mapsto q^i(t) p_i(t)$.
We find
\bes
\L(K_f) &=& \int dt \ \no{ ( - \Delta(f(t)) \dt q^i(t)
 - H^i_f(t) + \d_0f(t) q^i(t) ) p_i(t) } \nl
&&+ \Delta(f(t)) L(t) 
 + \d_0\Delta(f(t)) T^0_0(t) \\
&&+ (-)^i(-\d_0 H_f^i(t) + \d_0^2 f(t) q^i(t)) T^0_i(t)
 + (-)^{fj} \d_j\Delta(f(t)) T^j_0(t) \nl
&&+(-)^{(f+i)j+i} (-\d_jH_f^i(t) + \d_j\d_0f(t) q^i(t) 
 + \dlt^i_j\d_0f(t)) T^j_i(t),
\nonumber
\ees
where $f(t) = f(q(t))$, $H_f^i(t) = H_f^i(q(t))$.
The extension can be computed from (\ref{Kf}) and (\ref{ext}), but
the calculations are tedious and the result is not very illuminating.

Actually, there are two different notions of time in our representation of 
the contact algebra: one is treated specially in (\ref{Lxi}) and the other
in (\ref{KB}). Although it is natural to identify these two times, as we have
done above, this is not necessary. A more general representation of extended
$K(\N|M)$ is obtained as follows.
Introduce fixed constant vectors $z_\mu$ and $z^\mu$ with components only in
the bosonic directions, satisfying
\be
z_\mu z^\mu = 1, \qquad z_\mu \om^{\mu\nu} = 0.
\ee
The case above is recovered when $z_\mu = \dlt^0_\mu$, $z^\mu = \dlt^\mu_0$.
Then
\bes
E &=& x^\mu\dmu - z_\mu x^\mu z^\nu \dnu, \qquad \Delta = 2-E \nl
\{f, g\}_\KB &=& \Delta(f(x)) z^\mu\dmu g(x) - z^\mu\dmu f(x) \Delta(g(x)) -
\{f, g\}_\PB \nl
K_f &=& \Delta(f(x)) z^\mu\dmu - H_f + z^\mu\dmu f(x) E,
\ees
where the Poisson bracket is given by (\ref\PB) and $H_f$ by (\ref{Hf}).
The action in Fock space is given by
\bes
\L(K_f) &=& \int dt\ :\big\{ \Delta(f(t)) z^\mu 
 - (-)^{f\nu+\mu} \dnu f(t)\om^{\nu\mu} \nl
\bl+ z^\nu\dnu f(t) (q^\mu(t) - z_\si q^\si(t) z^\mu) \big\} p_\mu(t): \nl
&&+\big\{ \Delta(f(t)) z^0 - (-)^{f\nu} \dnu f(t)\om^{\nu0} \nl
 \bl+ z^\nu\dnu f(t)(t - z_\si q^\si(t) z^0)  \big\} L(t) \nl
&&+ (-)^{(f+\mu)\rho+\rho} \drho  \big\{ \Delta(f(t)) z^\mu - 
 (-)^{f\nu+\mu} \dnu f(t)\om^{\nu\mu} \nl
\bl+ z^\nu\dnu f(t) (q^\mu(t) - z_\si q^\si(t) z^\mu)  \big\} T^\rho_\mu(t),
\ees
where $f(t) \equiv f(q(t))$.

\subsection{ Superconformal algebra }     
\label{supconf}
As is known \cite{GLS97}, the superconformal algebra is a central extension 
of the contact algebra $K(1|1)$. 
Denote $t = x^0$ and $\th=x^1$, $\deg t = 0$, $\deg \th = 1$.
A Fourier basis for functions in $(1|1)$ dimensions is given by
\be
\ell_m = {1\/2i} \emt, \qquad 
g_m = \th  \emt.
\ee
Let $\om^{11} = i$ be the only non-zero component of the matrix
$\om^{\mu\nu}$.
The Poisson brackets are
\be
\{\th, \th\}_\PB = i, \qquad 
\{t, t\}_\PB = \{t, \th\}_\PB = 0.
\ee
With $\Delta = 2-\th\d_\th$, we have 
$\Delta(\ell_m) = 2\ell_m$, $\Delta(g_m) = g_m$.
The functions $\ell_m$ and $g_m$ generate a centerless superconformal algebra
under the contact bracket (\ref{KB}).
\bes
\{\ell_m, \ell_n\}_\KB &=& (n-m) \ell_{m+n} \nl
\{\ell_m, g_n\}_\KB &=& (n-{m\/2}) g_{m+n} \nl
\{g_m, g_n\}_\KB &=& 2 \ell_{m+n}
\ees
The corresponding contact vector fields are
\bes
K(\ell_m) &=& \emt( -i\d_t + {m\/2}\th\d_\th ) \nl
K(g_m) &=& \emt( \th\d_t -i\d_\th )
\ees
Substitution of these vector fields into (\ref{Lxi}) yields
\bes
L_m &\equiv& \L(K(\ell_m)) \nl
&=& \int dt\ \emt \Big\{ 
 i\no{\dt\th(t) p_\th(t)} + {m\/2}\no{\th(t)p_\th(t)} - iL(t) \nl
 &&+m T^0_0(t) -i{m^2\/2}\th(t) T^0_1(t) + {m\/2} T^1_1(t) \Big\} \nl
G_m &\equiv& \L(K(g_m)) \nl
&=& \int dt\ \emt \Big\{ 
 \no{ \dt\th(t)\th(t)p_\th(t)} - ip_\th(t) + \th(t)L(t) \nl
 &&+ im\th(t) T^0_0(t) - mT^0_1(t) - T^1_0(t) \Big\}.
\label{LG}
\ees
Introduce
\bes
\th_m &=& \nnni\int dt\ \th(t) \emt \nl
U_m &=& \nnni\int dt\ \dt\th(t)\th(t) \emt \nl
V_m &=& \nnni\int dt\ \ddt\th(t)\dt\th(t) \emt \nl
W_m &=& \nnn\int dt\ \ddt\th(t)\dt\th(t)\th(t) \emt.
\label{tUVW}
\ees
These operators satisfy the following Lie superalgebra.
\bes
[L_m,L_n]_- &=& (n-m)L_{m+n} + (-am^3 + a'm) \dlt_{m+n} \nl
{[}L_m,G_n]_- &=& (n-{m\/2}) G_{m+n} + (\al m^3+\beta m^2n) \th_{m+n} \nl
 &&- \gamma (mn+{m^2\/2}) \th_{m+n} - 2\gamma' m\th_{m+n} \nl
{[}G_m,G_n]_+ &=& 2L_{m+n} + (b m^2 + 2\gamma'-a') \dlt_{m+n} \nl
 &&+ ((2\al-e)(m^2+n^2) + (2\beta-e) mn) U_{m+n} + e V_{m+n} \nl
 &&-\gamma(m+n)U_{m+n} - 2\gamma' U_{m+n} \nl
{[}L_m,\th_n]_- &=& (n+{3m\/2})\th_{m+n} \nl
{[}L_m, U_n]_- &=& (n+m) U_{m+n} 
\label{xsc} \\
{[}L_m,V_n]_- &=& (n-m)V_{m+n} + (m^3+{m^2n\/2})U_{m+n} \nl
{[}L_m,W_n]_- &=& (n-{m\/2})W_{m+n} \nl
{[}G_m,\th_n]_+ &=& - \dlt_{m+n} + U_{m+n} \nl
{[}G_m,U_n]_- &=& (2m+n) \th_{m+n} \nl
{[}G_m,V_n]_- &=& (n-2m)W_{m+n} + (2m^3+3m^2n+mn^2) \th_{m+n} \nl
{[}G_m,W_n]_+ &=& V_{m+n} - (2m^2+mn) U_{m+n} \nl
{[}\th_m, \th_n]_+ &=& {[}\th_m, U_n]_- = {[}\th_m, V_n]_- 
= {[}\th_m, W_n]_+ = {[}U_m, U_n]_- = {[}U_m, V_n]_- \nl
&=& {[}U_m, W_n]_- = {[}V_m, V_n]_- = {[}V_m, W_n]_- = {[}W_m, W_n]_+ = 0,
\nonumber
\ees
where $2\al - \beta = 2a - b/2$.
The symmetry of the bracket has here been made explicit; $[\cdot,\cdot]_-$
is the commutator and $[\cdot,\cdot]_+$ the anti-commutator.
It is straightforward to verify all super-Jacobi identities and graded
anti-symmetry.
All parameters except $a$ can be removed by the following redefinition.
\bes
L_m &\mapsto& L_m - {a'\/2} \dlt_m + {\beta\/2} m^2 U_m, \nl
G_m &\mapsto& G_m - {e\/2} W_m + ((\beta-2\al)m^2 + \gamma m + \gamma')
\th_m.
\label{redef}
\ees
Eq. (\ref{xsc}) then contains a superconformal
subalgebra with central charge $12a$.
\bes
[L_m,L_n]_- &=& (n-m)L_{m+n} + (-am^3 + a'm) \dlt_{m+n} \nl
{[}L_m,G_n]_- &=& (n-{m\/2}) G_{m+n}  \nl
{[}G_m,G_n]_+ &=& 2L_{m+n} +(4a m^2 -a') \dlt_{m+n}.
\label{scalg}
\ees
The generators in (\ref{LG}) and (\ref{tUVW}) satisfy the algebra
(\ref{xsc}) with parameters
\bes
a &=& -{3\/4} + {3\/4}k_1 + {1\/4}k_2 + {c-2\/12},
\qquad a' = {c-2\/12}, \nl
\al &=& -{5\/4}+ {c-2\/12},
\qquad \beta = -{k_1\/2} - {k_2\/2} -1, \nl 
\gamma &=& -{1\/2}, 
\qquad \gamma' = {c-2\/24}, \nl
e &=& -2+ {c-2\/12},
\qquad b = 2k_1.
\ees
The central charge of (\ref{scalg}) is $12a = -11+9k_1+3k_2+c$.
Hence we have constructed a representation of the superconformal algebra
for each $Vir \ltimes gl(1|1)$ module.
It is possible to make the projection
$\th_m = 0$, $U_m = \dlt_m$ in (\ref{xsc}), because this choice makes
these operators central. A motivation comes from the Berezin integral:
$\int dt\ \dt\th(t)\th(t) \approx \int d\th \th = 1$.

\subsection{ $M=2$ superconformal algebra }
The superconformal algebra with two supersymmetries is a central extension 
of the contact algebra $K(1|2)$.
The coordinates are
$t = x^0, \th=x^1, \thb=x^2$, $\deg t = 0$, $\deg \th = \deg\thb = 1$.
A Fourier basis for functions in $(1|2)$ dimensions is given by
\bes
\ell_m = {1\/2i} \emt, &\qquad&
t_m = i\th\thb\emt \nl 
g_m = \th\emt, &\qquad&
\gb_m = \thb\emt.
\ees
Let $\om^{12} = i$ be the only non-zero component of the matrix
$\om^{\mu\nu}$.
The Poisson brackets are
\bes
\{\th, \thb\}_\PB &=& i, \nl
\{\th, \th\}_\PB &=& \{\thb, \thb\}_\PB = \{t, t\}_\PB 
 = \{t, \th\}_\PB =\{t, \thb\}_\PB = 0.
\ees
Moreover, $\Delta = 2-\th\d_\th - \thb\d_\thb$ and
\bes
\Delta(\ell_m) = 2\ell_m, &\qquad&
\Delta(t_m) = 0 \nl
\Delta(g_m) = g_m, &\qquad&
\Delta(\gb_m) = \gb_m.
\ees
The functions $\ell_m$, $g_m$, $\gb_m$ and $t_m$ generate a centerless 
$M=2$ superconformal algebra under the contact bracket (\ref{KB}).
\bes
\{\ell_m, \ell_n\}_\KB&=& (n-m) \ell_{m+n} \nl
\{\ell_m, t_n\}_\KB &=& n t_{m+n} \nl
\{\ell_m, g_n\}_\KB &=& (n-{m\/2}) g_{m+n} \qquad
\{\ell_m, \gb_n\}_\KB = (n-{m\/2}) \gb_{m+n} \nl
\{t_m, g_n\}_\KB &=& g_m, \qquad
\{t_m, \gb_n\}_\KB = -\gb_m, \nl
\{g_m, \gb_n\}_\KB &=& 2 \ell_{m+n} + (n-m)t_{m+n} \nl
\{g_m, g_n\}_\KB &=& \{\gb_m, \gb_n\}_\KB = \{t_m, t_n\}_\KB = 0.
\ees
The corresponding contact vector fields are
\bes
K(\ell_m) &=& \emt( -i\d_t + {m\/2}(\th\d_\th + \thb\d_\thb) ) \nl
K(g_m) &=& \emt( \th\d_t -i\d_\thb + im\th\thb\d_\thb ) \nl
K(\gb_m) &=& \emt( \thb\d_t -i\d_\th -im\th\thb\d_\th ) \nl
K(t_m) &=& \emt( \th\d_\th - \thb\d_\thb ).
\ees
Substitution of these vector fields into (\ref{Lxi}) yields a Fock 
representation of an extension of $K(1|2)$. The extension is readily 
found from (\ref{ext}). We have not explicitly calculated
it, since the calculation is quite tedious and the result
is not very illuminating. However, by analogy with the $M=1$ case,
we expect that the resulting non-central extension has a subalgebra
isomorphic to the standard centrally extended $M=2$ superconformal algebra.

\section{Gauge superalgebras}

\begin{theorem}\label{gaugethm} (cf. \cite{Lar97}, theorem 7.1)
Let $J^a(t)$ obey the Kac-Moody superalgebra $\km\oj$ and let $q^\mu(t)$
be a trajectory (\ref{traj}). Then
\be
\J_X = \int dt\ X_a(q(t)) J^a(t)
\label{JX}
\ee
satisfies a Lie superalgebra extension of $map(\N|M,\oj)$, the algebra of
maps from $(\N|M)$-dimensional super space-time to $\oj$.
The brackets are
\bes
[\J_X, \J_Y] &=& \J_{[X,Y]} 
 - k \int \dx\ (-)^{a(Y+b)} S_1^\mu(x) \dmu X_a(x) Y_a(x) \dlt^{ab} \nl
{[}\J_X, S_1^\nu(x)] &=& 0,
\label{extJ}
\ees
where 
\be
[X,Y]_c = (-)^{a(Y+b)} if^{ab}{}_c X_a Y_b = -(-)^{XY} [Y,X]_c.
\ee
The Killing super-metric $\dlt^{ab}$ was defined in (\ref{Killing}), and
\be
\int \dx\ S_1^\nu(x) h_\nu(x) = \nnni\int dt\ \qnu(t) h_\nu(q(t)),
\ee
as in (\ref{SRq}) and (\ref{SR}).
Moreover, there is an intertwining action of $diff(\N|M)$:
\bes
[\Lxi, \J_Y] &=& \J_{\xi Y}, \\
{[}\Lxi, S_1^\nu(x)] &=&
 -\xmu(x)\dmu S_1^\nu(x) - \div\xi(x) S_1^\nu(x) 
 + (-)^{\mu(\xi+\mu+\nu)} \dmu\xi^\nu(x) S_1^\mu(x),
\nonumber
\ees
where $\xi Y = \xmu(x) \dmu Y(x)$.
\end{theorem}

\section{Discussion}

To conclude, I have discovered a class of non-central extensions of the
super-diffeomorphism and super-gauge algebras in any number of bosonic
and fermionic dimensions, and constructed Fock representations thereof.
To my knowledge this is the first time non-central extensions of
Lie superalgebras have been described. 

The superconformal algebra was believed to be an exceptional algebraic
structure, because it is one of the few superalgebras admitting a central
extension (an exhaustive list is given in \cite{GLS97}). However, the 
interesting property, both mathematically and physically, is that an algebra
has projective Fock representations, not that the resulting extension 
necessarily be central. 
In subsection \ref{supconf} I proved that the superconformal algebra
is not exceptional at all, but rather a quite ordinary
(and indeed one of the simplest) subalgebras of the non-centrally extended
super-diffeomorphism algebra.

This result has some bearing on string theory. Since there is no compelling
experimental evidence in favour of string theory, the only motivation has
been that it represents an exceptional mathematical structure. 
However, the present paper shows that there is nothing special 
about the underlying algebraic structure, the superconformal algebra, 
and hence I find it difficult to believe
that string theory should be relevant to physics. Stated differently: the 
superconformal algebra is merely a subalgebra of the non-centrally
extended super-diffeomorphism algebra in the lowest possible dimension.
It seems unlikely that Nature should prefer such a trivial symmetry
at her most fundamental level.

Another motivation for string theory is that it may provide a consistent
theory of quantum gravity. Fortunately, $diff(\N)$ can achieve the
same goal, in the following sense. Its Fock modules are projectively
generally covariant (the gravity property), they are Fock spaces with
energy bounded from below (the quantum property), and all matrix elements
of normal ordered operators are manifestly finite (the consistency
property). Moreover, I showed in \cite{Lar97} that trajectory analogs of
the Einstein and geodesic equations are well defined, thus making the
connection to Einsteinian gravity closer.

Finally, there is also a philosophical motivation for studying space-time 
diffeomorphisms. Namely, from a passive point of view, a diffeomorphism is
simply a coordinate transformation. Because every physical object must be
invariantly defined, rather than being an artifact of the choice of
coordinate
system, it must transform consistently under arbitrary space-time
diffeomorphisms,
i.e. as a representation. This also contains information
about the dynamics, since the Hamiltonian is itself a space-time
diffeomorphism;
it is the generator of rigid time translations. Thus, the $diff(\N)$
representation theory amounts to a classification of all physical objects. 
A list of unitary irreps is urgently needed.

\end{document}